\documentclass[conference]{IEEEtran}
\usepackage{cite}
\usepackage{amsmath,amssymb,amsfonts,amsthm}
\usepackage{algorithmic}
\usepackage{graphicx}
\usepackage{stfloats}
\usepackage{lipsum}
\usepackage{float}
\usepackage{textcomp}
\usepackage{steinmetz}
\usepackage[utf8]{inputenc}
\usepackage[english]{babel}
\newtheorem{theorem}{Theorem}

\makeatletter
\usepackage{array,booktabs,ragged2e}
\newcolumntype{R}[1]{>{\RaggedLeft\arraybackslash}p{#1}}
\def\BibTeX{{\rm B\kern-.05em{\sc i\kern-.025em b}\kern-.08em
    T\kern-.1667em\lower.7ex\hbox{E}\kern-.125emX}}
\begin{document}
\title{Synthesizing Averaged Virtual Oscillator Dynamics to Control Inverters with an Output LCL Filter}
\thanks{The corresponding author is Mr. Muhammad Ali. This research work has not been presented at any conference or submitted for publication elsewhere and is the original work of the authors.}
\author{\IEEEauthorblockN{}M. Ali$^*$, H. I. Nurdin and J. E. Fletcher\\
\textit{School of Electrical Engineering and Telecommunications, University of New South Wales, Australia}\\
%\textit{Deparment of Energy Technology, Aalborg University, Esbjerg, Denmark.}\\
$^*$m.ali@unsw.edu.au\\
}
%\vspace{-1cm}
%m.ali@unsw.edu.au}
%\and
%\IEEEauthorblockN{Hendra I. Nurdin}
%\IEEEauthorblockA{\textit{School of Electrical Engineering and Telecommunications} \\
%\textit{University of New South Wales}\\
%Sydney, NSW 2052, Australia\\
%h.nurdin@unsw.edu.au}
%\and
%\IEEEauthorblockN{John Fletcher}
%\IEEEauthorblockA{\textit{School of Electrical Engineering and Telecommunications} \\
%\textit{University of New South Wales}\\
%Sydney, NSW 2052, Australia\\
%john.fletcher@unsw.edu.au}
%}
%
\maketitle
\begin{abstract}
In commercial inverters, an LCL filter is considered an integral part to filter out the switching harmonics and generate a sinusoidal output voltage. The existing literature on the averaged virtual oscillator controller (VOC) dynamics is for current feedback before the output LCL filter that contains the switching harmonics or for inductive filters ignoring the effect of filter capacitance. In this work, a new version of averaged VOC dynamics is presented for islanded inverters with current feedback after the LCL filter thus avoiding the switching harmonics going into the VOC. The embedded droop-characteristics within the averaged VOC dynamics are identified and a parameter design procedure is presented to regulate the output voltage magnitude and frequency according to the desired ac-performance specifications. Further, a power dispatch technique based on this newer version of averaged VOC dynamics is presented to simultaneously regulate both the active and reactive output power of two parallel-connected islanded inverters. The control laws are derived and a power security constraint is presented to determine the achievable power set-point. Simulation results for load transients and power dispatch validate the proposed version of averaged VOC dynamics.\\
\end{abstract}
\begin{IEEEkeywords}
Virtual oscillator control, averaged model, LCL filter, inverter control, power dispatch, security constraint.
\end{IEEEkeywords}
\section{Introduction}
Power systems are going through a transitional period where the conventional synchronous generators are being replaced by the renewable energy sources (RESs). The rapid integration of RESs into the power system has resulted in a need to investigate control techniques to stabilise and regulate the voltage and frequency for systems with low or zero inertia. Existing techniques for systems with RESs include droop control \cite{de2007voltage,mohamed2008adaptive}, proportional resonant (PR) control \cite{teodorescu2004proportional,shoeiby2014voltage} and recently proposed virtual oscillator control (VOC) \cite{johnson2016synthesizing,johnson2014synchronization,ali2019power}.

Virtual oscillator is a nonlinear Van der Pol oscillator that exhibits nearly sinusoidal oscillations in the steady-state. In contrast to existing inverter control techniques, virtual oscillator controller only requires the inverter output current as a feedback and regulates the output voltage (magnitude and frequency) according to the desired ac-performance specifications. In \cite{johnson2016synthesizing}, an averaged VOC model is derived to explicitly identify the $P-V$ and $Q-\omega$ droop-characteristics embedded within the averaged dynamics of the virtual oscillator controller. Recent literature on the dispatchability of virtual oscillator controlled inverters can be found in \cite{ali2018output,ali2018simultaneous,ali2019regulation,gross2019effect,raisz2018power,seo2019dispatchable,ali2020TIE}.     

The averaged VOC model presented in \cite{johnson2016synthesizing} is derived when there is a current feedback to the controller before the output filter. However, this model is not an accurate representation when there is a current feedback after the output LCL filter. In case of commercial inverters, an output LCL (or LC) filter is always considered an essential part to filter out the switching harmonics and improve the quality of output voltage. In this work, a new version of averaged VOC dynamics is presented that takes into account the LCL filter at the output of an inverter. In the proposed control scheme, the current is feedback after the LCL filter thus avoiding the switching harmonics going into the virtual oscillator controller. The contribution of this work is twofold. At first, an averaged VOC model is derived for an inverter with current feedback after the output LCL filter. The updated embedded droop-characteristics within the proposed averaged VOC dynamics are presented and corresponding equilibrium points are derived. A system's parameter design procedure has been presented to regulate the inverter's output voltage magnitude and frequency according to the desired ac-performance specifications. In order to validate the proposed averaged VOC dynamics, simulation results are presented for a number of scenarios including rise time, harmonics analysis, droop characteristics and load transients. It has been shown that for all the scenarios, the proposed averaged VOC model predicts the actual VOC (Van der Pol) dynamics accurately. A comparison with the previously reported averaged VOC model \cite{johnson2016synthesizing} is also presented. The previously derived averaged VOC dynamics \cite{johnson2016synthesizing} starts to differ more from the actual VOC dynamics (with current feedback after the LCL filter) for a large value of filter capacitance. This is due to the fact that a higher value of filter capacitance draws more current and results in a significant difference between the current before and after the filter.

The second contribution is the extension of power dispatch technique \cite{ali2018simultaneous} to this new version of averaged VOC dynamics. The updated control laws are derived based on this new model to determine the control inputs corresponding to a particular power set-point. Further, an updated power security constraint is derived to determine the feasible power set-points. Using this constraint, it can be determined a-priori if a particular power set-point is achievable or not. Simulation results are presented for a number of power dispatch scenarios to validate the proposed power dispatch technique.

The rest of the paper is organised as follows. In Section \ref{sec:System Description}, an overall system description is presented. In Section \ref{sec:System Modelling}, an averaged VOC model is derived for inverters with current feedback after the LCL filter. In Section \ref{sec:VOC Parameters Design Procedure}, a VOC parameter design procedure is discussed. In Section \ref{sec:Power Dispatch}, a power dispatch technique is presented. In Section \ref{sec:Simulation Results}, simulation results are presented. In Section \ref{sec:Conclusion}, conclusion is drawn.
\section{System Description}
\label{sec:System Description}
\begin{figure}[t]
\centering
\includegraphics[width=0.9\columnwidth]{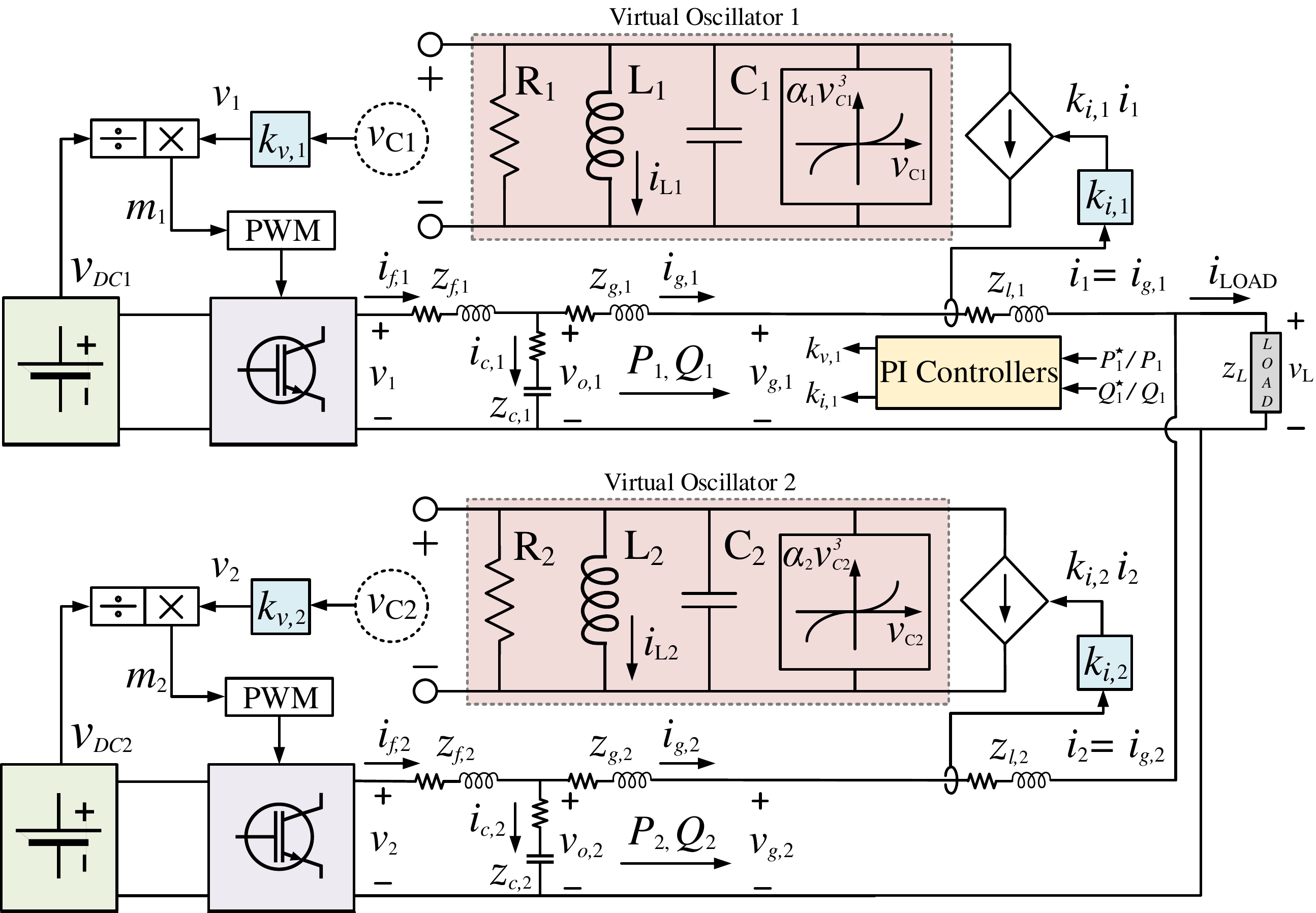}\\%{drivingpeaksrsq}
\vspace{-0.2 cm}
    \caption{A overview of the overall system consisting of two parallel-connected virtual oscillator controlled inverters. The inverter $1$ can dispatch power using the PI controllers while inverter $2$ supplies the remaining power.}
\label{fig:twovocsystem}
\vspace{-0.5 cm}
\end{figure}

The system considered consists of two virtual oscillator controlled inverters with current feedback after the output LCL filter. A complete system overview is shown in Fig. \ref{fig:twovocsystem}. The filter parameters are $z_f = R_f + j\omega^*L_f$, $z_c = R_c + \frac{1}{j\omega^*C_f}$ and $z_g = R_g + j\omega^*L_g$. The line impedance is represented as $z_l = R_l + j\omega^*L_l$. The inverters are connected to a common RL load $z_{L} = R_{L} + j\omega^*L_{L}$ through the point of common coupling (PCC). Moreover, inverter $1$ can dispatch power by the use of two additional PI controllers.
\section{System Modelling}
\label{sec:System Modelling}
In this section, the VOC dynamics are presented. Further, an averaged VOC model is derived for inverters with current feedback after the output LCL filter. The dynamics of the PI controllers used for power dispatch are the same as in \cite{ali2018simultaneous}.
%
%\vspace{-0.4 cm}
\subsection{Virtual Oscillator Controller}
\label{subsec:Virtual Oscillator Controller}
A virtual oscillator consists of an LC harmonic oscillator with a fundamental frequency $\omega^* = 1/\sqrt{LC}$. It consists of a negative resistance element $R = -1/\sigma$ and a nonlinear current source with a positive constant $\alpha$ as shown in Fig. \ref{fig:twovocsystem}. The feedback current $i$ is scaled by the current feedback gain $k_i$ before entering the virtual oscillator. The $v_C$ denotes the capacitor voltage and $i_L$ denotes the inductor current. The capacitor voltage $v_C$ is scaled by the voltage scaling factor $k_v$ to generate the inverter output voltage $v$. The actual VOC dynamics are given by the following dynamic equations \cite{johnson2016synthesizing}:
\small
\begin{align}\label{eq:voc actual model}
    \notag
    \frac{dV}{dt} &= \frac{\epsilon\omega^*}{\sqrt{2}} \left( \sigma g\big(\sqrt{2}V\cos{(\phi)}\big) - k_{v}k_{i}i\right)\cos(\phi),\\
    \frac{d\phi}{dt} &= \omega^* - \frac{\epsilon\omega^*}{\sqrt{2}V}\left(\sigma g\big(\sqrt{2}V\cos(\phi)\big)-k_{v}k_{i}i\right)\sin(\phi).
\end{align}
\normalsize
A detailed derivation and parametric description of the dynamical model for the inverter RMS output voltage magnitude $V$ and instantaneous phase angle $\phi = \omega t + \theta$ where $\theta$ is the phase offset with respect to $\omega t$, can be found in \cite{johnson2016synthesizing}.
\subsection{Averaged VOC Model for Inverters with Output LCL Filter}
\label{subsec:Averaged VOC Model for Inverters with an Output LCL Filter}
A new version of averaged VOC dynamics is presented for inverters with current feedback after the output LCL filter. A brief derivation is in Appendix \ref{subsec:Derivation of the Averaged VOC Model with an LCL Filter}. The proposed averaged VOC model is given by the following dynamic equations:
\footnotesize
\begin{align}\label{eq:voc actual model v}
    \frac{d}{dt}\overline{V} = \frac{\sigma}{2C}\left( \overline{V} - \frac{\beta}{2}\overline{V}^3\right) - \frac{k_{v}k_{i}}{2C} \left ( \frac{C_{\alpha} \overline{P}}{\overline{V}} + \frac{S_{\alpha} \overline{Q}}{\overline{V}} + C_{\beta}\overline{V} \right ),
\end{align}
\footnotesize
\begin{align}\label{eq:voc actual model t}
    \frac{d}{dt}\overline{\theta} = \omega^* - \omega + \frac{k_{v}k_{i}}{2C} \left ( \frac{C_{\alpha} \overline{Q}}{\overline{V}^2} - \frac{S_{\alpha} \overline{P}}{\overline{V}^2} - S_{\beta} \right ),
\end{align}
\normalsize
where $\beta = \frac{3\alpha}{k_v^{2}\sigma}$, $\overline{V}$ denotes the averaged RMS output voltage magnitude and $\overline{\theta}$ denotes the averaged phase offset with respect to $\omega t$. The averaged active and reactive power are denoted by $\overline{P}$ and $\overline{Q}$, respectively. $C_{\alpha}, C_{\beta}, S_{\alpha}$ and $S_{\beta}$ are the impedance constants (as defined in Appendix \ref{subsec:Derivation of the Averaged VOC Model with an LCL Filter}). 
\subsubsection{Voltage Regulation Characteristics}
\label{subsubsec:Voltage Regulation Characteristics}
The equilibrium value for the averaged RMS voltage magnitude $\overline{V}_{eq}$ can be determined by setting the left hand side of \eqref{eq:voc actual model v} to zero and solving for $\overline{V}$ as follows:
\small
\begin{align}\label{eq:equil V1}
    \overline{V}_{eq} = k_v \sqrt{\frac{ \sigma_{\beta} \pm \sqrt{\sigma_{\beta}^2 - 6 \alpha \left ( k_i/k_v \right) \left ( C_{\alpha} \overline{P}_{eq} + S_{\alpha} \overline{Q}_{eq} \right ) }}{3\alpha}},
\end{align}
\normalsize
where $\sigma_{\beta} = \left ( \sigma - k_v k_i C_{\beta} \right )$. The  $\overline{P}_{eq}$ and $\overline{Q}_{eq}$ denote the active and reactive power at the equilibrium. Both the roots of \eqref{eq:equil V1} are real valued if the following inequality holds:
\small
\begin{align}\label{eq:equil V2}
    \left ( C_{\alpha} \overline{P} + S_{\alpha} \overline{Q} \right ) < S_{cr} := \frac{\sigma_{\beta}^2}{6 \alpha  \left ( k_i/k_v \right) },
\end{align}
\normalsize
where $k_i > 0$ and $k_v > 0$. $S_{cr}$ denotes the critical power constant with corresponding critical value of the inverter RMS output voltage $\overline{V}_{cr} = k_v \sqrt{\frac{\sigma_{\beta}}{3 \alpha}}$.
A local stability analysis of the high voltage solution of \eqref{eq:equil V1} (similar to \cite{johnson2016synthesizing}) is not considered. In the subsequent analysis, we assume that the high voltage solution of \eqref{eq:equil V1} is locally asymptotically stable and with a slight abuse of notation, we  denote it by $\overline{V}_{eq}$. Using \eqref{eq:equil V1}, the open circuit voltage $\overline{V}_{oc} = k_v \sqrt{\frac{2 \sigma_{\beta}}{3 \alpha}}$ for the VO-controlled inverter (i.e. $\overline{P}_{eq} = \overline{Q}_{eq} = 0$).
\subsubsection{Frequency Regulation Characteristics}
\label{subsubsec:Frequency Regulation Characteristics}
Solving the phase angle dynamics \eqref{eq:voc actual model t} for equilibrium point gives the system's frequency $\omega_{eq}$ in steady-state as follows:
\small
\begin{align}\label{eq:equil V4}
    \omega_{eq} = \omega^* + \frac{k_{v}k_{i}}{2C} \left ( \frac{C_{\alpha} \overline{Q}_{eq}}{\overline{V}^2_{eq}} - \frac{S_{\alpha} \overline{P}_{eq}}{\overline{V}^2_{eq}} - S_{\beta} \right ),
\end{align}
\normalsize
where $\overline{V}_{eq}$ is the high voltage solution of \eqref{eq:equil V1}.
\subsubsection{Dynamic Response}
\label{subsubsec:Dynamic Response}
In order to quantify the dynamic response of the VO-controlled inverter, the time taken by the inverter to reach its open circuit voltage $\overline{V}_{oc}$ is considered. The following voltage dynamics of interest in a variable-separable ODE form are obtained from \eqref{eq:voc actual model v} by replacing $\overline{P} = \overline{Q} = 0$:
\small
\begin{align}\label{eq:equil V5}
    \frac{d}{dt}\overline{V} = \frac{\sigma}{2C}\left( \overline{V} - \frac{\beta}{2}\overline{V}^3\right) & - \frac{k_{v}k_{i}C_{\beta}}{2C} \overline{V}.
\end{align}
\normalsize
The rise time $t_{rise}$ is the time taken by the inverter to build-up output voltage from $0.1\overline{V}_{oc}$ to $0.9\overline{V}_{oc}$. The $t_{rise} \approx \frac{6}{\omega^* \epsilon \sigma_{\beta}}$ is determined by separating variables in \eqref{eq:equil V5} and integrating under the limits from $0.1\overline{V}_{oc}$ to $0.9\overline{V}_{oc}$, where $\epsilon = \sqrt{L/C}$.
\subsection{PI Controller Dynamics}
\label{subsec:PI Controller Dynamics}
The PI controller dynamics used to regulate active power are $k_v = K_P^p(\overline{P} - P^*) + e_p$ where $\dot{e}_p = K_I^p(\overline{P} - P^*)$. Similarly, the PI controller dynamics used to regulate reactive power are $k_i = K_P^q(\overline{Q} - Q^*) + e_q$ where $\dot{e}_q = K_I^q(\overline{Q} - Q^*)$. The $e_p$ and $e_q$ are the PI controllers' states \cite{ali2018simultaneous}.
\section{VOC Parameter Design Procedure}
\label{sec:VOC Parameters Design Procedure}
A parameter design procedure is presented for VO-controlled inverters with current feedback after the LCL filter. The parameters are selected such that the VO-controlled inverter satisfy the desired ac-performance specifications. 
\subsection{Designing the Scaling Factors}
\label{subsec:Designing the Scaling Factors}
The voltage scaling factor $k_v$ is chosen equal to $\overline{V}_{oc}$ to standardise the design procedure such that the virtual oscillator capacitor voltage is equal to $1$ V RMS when the inverter's output voltage is equal to $\overline{V}_{oc}$. The current feedback gain $k_i$ is chosen as the ratio of $\overline{V}_{min}$ to $\overline{S}_{max}$. The $\overline{V}_{min}$ corresponds to the constant $\overline{S}_{max}$ defined by:
\small
\begin{align}\label{eq:equil V8}
    \overline{S}_{max} = \max_{\overline{P}^2 + \overline{Q}^2 \le |\overline{S}_{rated}|^2} \left ( C_{\alpha} \overline{P} + S_{\alpha} \overline{Q} \right ).
\end{align}
\normalsize  
By choosing the gains as:
\small
\begin{align}\label{eq:equil V9}
    k_v := \overline{V}_{oc}, \quad \quad k_i : = \frac{\overline{V}_{min}}{\overline{S}_{max}},
\end{align}
\normalsize
the parallel-connected inverters in a system share the power proportional to their power ratings \cite{johnson2016synthesizing,johnson2014synchronization}.
\subsection{Designing the Voltage Regulation Parameters}
\label{subsec:Designing the Voltage Regulation Parameters}
A design procedure for the virtual oscillator negative resistance element $R = \frac{-1}{\sigma}$ and the coefficient of nonlinear current source $\alpha$ is presented. The proposed design procedure ensures the RMS output voltage to stay within the range: $\overline{V}_{min} \le \overline{V}_{eq} \le \overline{V}_{oc}$ for $\overline{S}_{max} \ge \left ( C_{\alpha} \overline{P}_{eq} + S_{\alpha} \overline{Q}_{eq} \right ) \ge 0$. The definition of $V_{oc}$ and the choice of $k_v$ in \eqref{eq:equil V9} results in $\sigma = \frac{3\alpha}{2} + k_i k_v C_{\beta}$.
Substituting $\left ( C_{\alpha} \overline{P}_{eq} + S_{\alpha} \overline{Q}_{eq} \right ) = \overline{S}_{max}$, $\overline{V}_{eq} = \overline{V}_{min}$, $k_v$ and $k_i$ as in \eqref{eq:equil V9}, $\alpha = \frac{2}{3}\sigma_{\beta}$ and $\sigma_{\beta} = \left ( \sigma - k_v k_i C_{\beta} \right )$ in the high voltage solution of \eqref{eq:equil V1}, we get:
 \small
\begin{align}\label{eq:equil V14}
    \sigma = \frac{\overline{V}_{oc}}{\overline{V}_{min}} \frac{\overline{V}_{oc}^2}{\overline{V}_{oc}^2 - \overline{V}_{min}^2} + \frac{\overline{V}_{min}\overline{V}_{oc}C_{\beta}}{\overline{S}_{max}}.
\end{align}
\normalsize
\subsection{Designing the Harmonic Oscillator Paramters}
\label{subsec:Designing the Harmonic Oscillator Paramters}
In order to derive a set of constraints to determine the parameters $L$ and $C$, the frequency regulation characteristics \eqref{eq:equil V4}, the rise time $t_{rise}$ and the ratio of the amplitude of the third harmonic to the fundamental \cite[Eq. 41]{johnson2016synthesizing} are considered. While designing the harmonic oscillator parameters, one of the design input is the maximum permissible frequency deviation $|\Delta \omega|_{max}$. Let us start with the frequency regulation characteristics in \eqref{eq:equil V4} and define the following constant:
\small
\begin{align}\label{eq:equil V15}
    \overline{S}_{|\Delta \omega|_{max}} = \max_{\overline{P}^2 + \overline{Q}^2 \le |\overline{S}_{rated}|^2} \left ( C_{\alpha} \overline{Q} - S_{\alpha} \overline{P} \right ).
\end{align}
\normalsize
Using the worst-case operating condition for the output voltage (corresponding to $\overline{S}_{max}$ that results in the minimum permissible voltage $\overline{V}_{min})$ and substituting the scaling factors from \eqref{eq:equil V9} into \eqref{eq:equil V4}, the lower bound on capacitance $C$ is given by:
\footnotesize
\begin{align}\label{eq:equil V16}
    C \ge \frac{1}{2|\Delta \omega|_{max}} \left( \frac{\overline{S}_{|\Delta \omega|_{max}}\overline{V}_{oc}}{\overline{V}_{min}\overline{S}_{max}} - \frac{S_{\beta}\overline{V}_{oc}\overline{V}_{min}}{\overline{S}_{max}} \right) =: C_{|\Delta \omega|_{max}}^{min}.
\end{align}
\normalsize 
Define the maximum permissible rise time $t_{rise}^{max}$ as the design input. Now, considering the $t_{rise}$ of an unloaded inverter and \eqref{eq:equil V14}, the upper bound on the capacitance $C$ is defined as:
\small
\begin{align}\label{eq:equil V17}
    C \le \frac{t_{rise}^{max}}{6} \frac{\overline{V}_{oc}}{\overline{V}_{min}} \frac{\overline{V}_{oc}^2}{\overline{V}_{oc}^2 - \overline{V}_{min}^2} =: C_{t_{rise}^{max}}.
\end{align}
\normalsize 
Finally, the third design input is the maximum-permissible ratio of the amplitude of third harmonic to the fundamental $\delta_{3:1}^{max}$ where $\delta_{3:1} = \frac{\epsilon \sigma}{8}$ as defined in \cite[Eq. 41]{johnson2016synthesizing}. Replacing \eqref{eq:equil V14} in the expression for $\delta_{3:1}^{max}$, we get another lower bound on the capacitance $C$ given by:
 \footnotesize
\begin{align}\label{eq:equil V18}
    C \ge \left ( \frac{1}{8 \omega^* \delta_{3:1}^{max}} \right ) \left( \frac{\overline{V}_{oc}}{\overline{V}_{min}} \frac{\overline{V}_{oc}^2}{\overline{V}_{oc}^2 - \overline{V}_{min}^2} + \frac{\overline{V}_{min}\overline{V}_{oc}C_{\beta}}{\overline{S}_{max}} \right) =: C_{\delta_{3:1}}^{min}.
\end{align}
\normalsize
The constraints \eqref{eq:equil V16}-\eqref{eq:equil V18} define a permissible range for the capacitance satisfying the desired frequency regulation, rise time and harmonic distortion specifications. The permissible range of capacitance $C$ can be written as:
 \small
\begin{align}\label{eq:equil V19}
    max\{ C_{|\Delta \omega|_{max}}^{min}, C_{\delta_{3:1}}^{min} \} \le C \le C_{t_{rise}^{max}}.
\end{align}
\normalsize 
Once the value is chosen for the capacitance $C$, the inductance can be determined as $L = \frac{1}{C(\omega^*)^2}$. Note that it may be possible \eqref{eq:equil V19} does not hold for the set of design inputs $\{ |\Delta \omega|_{max}, t_{rise}^{max}, \delta_{3:1}^{max}\}$ and necessitates a design trade-off.
\section{Power Dispatch}
\label{sec:Power Dispatch}
The power dispatch technique presented in \cite{ali2018simultaneous} is extended to the averaged VOC model with current feedback after the output LCL filter. Inverter $1$ simultaneously regulates both the active and reactive power according to the desired power set-point $(P_1^*, Q_1^*)$ using the PI controllers that continuously tune the VOC parameters $k_{v,1}$ and $k_{i,1}$ as in \cite{ali2018simultaneous}.
\subsection{Power Security Constraint}
\label{subsec:Power Security Constraint}
An updated power security constraint is derived to determine the achievable power set-points and guarantee the existence of real valued control inputs.
\begin{theorem}\label{thrm:t1}
Assuming the averaged model of two VOC inverters with current feedback after the output LCL filter that synchronise to a common frequency, and are connected to a common fixed impedance load $z_L$ through line impedance values $z_{l,1}$ and $z_{l,2}$, respectively, the desired output power set-point $\overline{P}_1^{*}$ and $\overline{Q}_1^{*}$ for inverter $1$ can be achieved and there exists corresponding real-valued current feedback gain $k_{i,1}^{\iota}$ and voltage scaling factor $k_{v,1}^{\iota}$, if the following security constraint is satisfied for the averaged VOC-dynamics:
\small
\begin{align}\label{eq:constraint}
    \sigma_1 {\overline{V}_1^{\iota}}^2 - \mu \left ( C_{\alpha,1} \overline{P}_1^{*} + S_{\alpha,1} \overline{Q}_1^{*} + C_{\beta,1} {\overline{V}_1^{\iota}}^2 \right) > 0,
\end{align}
\normalsize
where,
\small
\begin{align}\label{eq:gamma}
    \mu = k_{v,1}^{\iota} k_{i,1}^{\iota} = k_{v,2} k_{i,2} \frac{\left( \frac{C_{\alpha,2} \overline{Q}_2^{\iota} - S_{\alpha,2} \overline{P}_2^{\iota} }{{\overline{V}_2^{\iota}}^2} - S_{\beta,2} \right)}{\left( \frac{C_{\alpha,1} \overline{Q}_1^{*} - S_{\alpha,1} \overline{P}_1^{*} }{{\overline{V}_1^{\iota}}^2} - S_{\beta,1} \right)}.
\end{align}
\normalsize
$\overline{V}_2^{\iota}$ and $\overline{Q}_2^{\iota}$ denote the steady-state averaged output voltage magnitude and reactive power supplied by inverter $2$, respectively, corresponding to the desired power set-point ($\overline{P}_1^{*}, \overline{Q}_1^{*}$).
\end{theorem}
The proof of this theorem is similar to \cite[Theorem 1]{ali2018simultaneous} and is omitted due to the space limitation. The updated control laws to determine the control inputs $k_{v,1}^{\iota}$ and $k_{i,1}^{\iota}$ corresponding to the desired power set-point $(\overline{P}_1^{*}, \overline{Q}_1^{*})$ are:
\footnotesize
\begin{align}\label{eq:gamma11}
    k_{v,1}^{\iota} = \pm \sqrt{\frac{\left( \sigma_1  - \mu C_{\beta,1} \right) {\overline{V}_1^{\iota}}^4}{\left( \sigma_1  - \mu C_{\beta,1} \right){\overline{V}_1^{\iota}}^2 - \mu \left( C_{\alpha,1} \overline{P}_1^{*} + S_{\alpha,1} \overline{Q}_1^{*} \right)}}, \, k_{i,1}^{\iota} = \frac{\mu}{k_{v,1}^{\iota}}.
\end{align}
\normalsize
\section{Simulation Results}
\label{sec:Simulation Results}
Simulation results in MATLAB (with ideal voltage sources) are presented for a number of scenarios to validate the proposed averaged model. The line, load and filter parameters $R_f = R_l = 0.15 \, \Omega$, $L_f = L_l = 2.48$ mH, $R_c = 3.3 \, \Omega$, $C_f = 4.7 \, \mu$F, $R_g = 0.13 \, \Omega$, $L_g = 0.97$ mH, $R_L = 22.1 \, \Omega$ and $L_L = 14.4$ mH. The system parameters and ac-performance specifications are as in Table \ref{tab:System Parameters} and Table \ref{tab:ac specs}.    
\begin{table}[t]
\centering
\renewcommand{\arraystretch}{1.3}
\caption{System Parameters}
\label{tab:System Parameters}
\resizebox{0.85\columnwidth}{!}{
\begin{tabular}{p{0.7cm}p{3.7cm}ccc}
    \hline
    \hline
        Symbol & Parameter & Inverter $1$ & Inverter $2$ & Unit \\
    \hline
        $k_i$ & Current feedback gain & $0.15225$ & $0.15225$ & A/A \\
        $k_v$ & Voltage scaling factor & 126 & $126$  & V/V \\
        $\sigma$ & Conductance & $6.09256$ & $6.09256$ & $\Omega^{-1}$ \\
        $\alpha$ & Cubic-current source coefficient & $4.06184$ & $4.06184$ & A/V$^3$ \\
        $L$ & Oscillator inductance & $34.661$ & $34.661$ & $\mu$H \\
        $C$ & Oscillator capacitance & $0.203$ & $0.203$ & F \\
    \hline		
    \hline
\end{tabular}}
\vspace{-0.3 cm}
\end{table}
\vspace{-0.0 cm}
\begin{table}[t]
\centering
\renewcommand{\arraystretch}{1.3}
\caption{AC Performance Specifications}
\label{tab:ac specs}
\tiny
\resizebox{0.85\columnwidth}{!}{
\begin{tabular}{p{0.7cm}p{2.05cm}cc}
    \hline
    \hline
        Symbol & Parameter & Value & Unit \\
    \hline
        $\overline{V}_{oc}$ & RMS open-circuit voltage & $126$ & V \\
        $\overline{V}_{min}$ & RMS rated power voltage & $114$ & V \\
        $\overline{P}_{rated}$ & Rated real power & $750$ & W \\
        $|\overline{Q}_{rated}|$ & Rated reactive power & $750$ & VAr\\
        $\omega^*$ & Nominal oscillator frequency & $2\pi60$ & rad/s \\
        $|\Delta \omega|_{max}$ & Maximum frequency offset & $2\pi0.5$ & rad/s \\
        $t_{rise}^{max}$ & Maximum rise time & $0.2$ & s \\
        $\delta_{3:1}^{max}$ & $3^{\mbox{rd}}$ to $1^{\mbox{st}}$ harmonic ratio & $1$ & $\%$ \\
    \hline
    \hline
\end{tabular}}
\vspace{-0.5 cm}
\end{table}
\vspace{-0.1 cm} 
\subsection{Embedded Droop Characteristics}
\label{subsec:Embedded Droop Characteristics}
\begin{figure}[t]
\centering
\includegraphics[width=0.9\columnwidth]{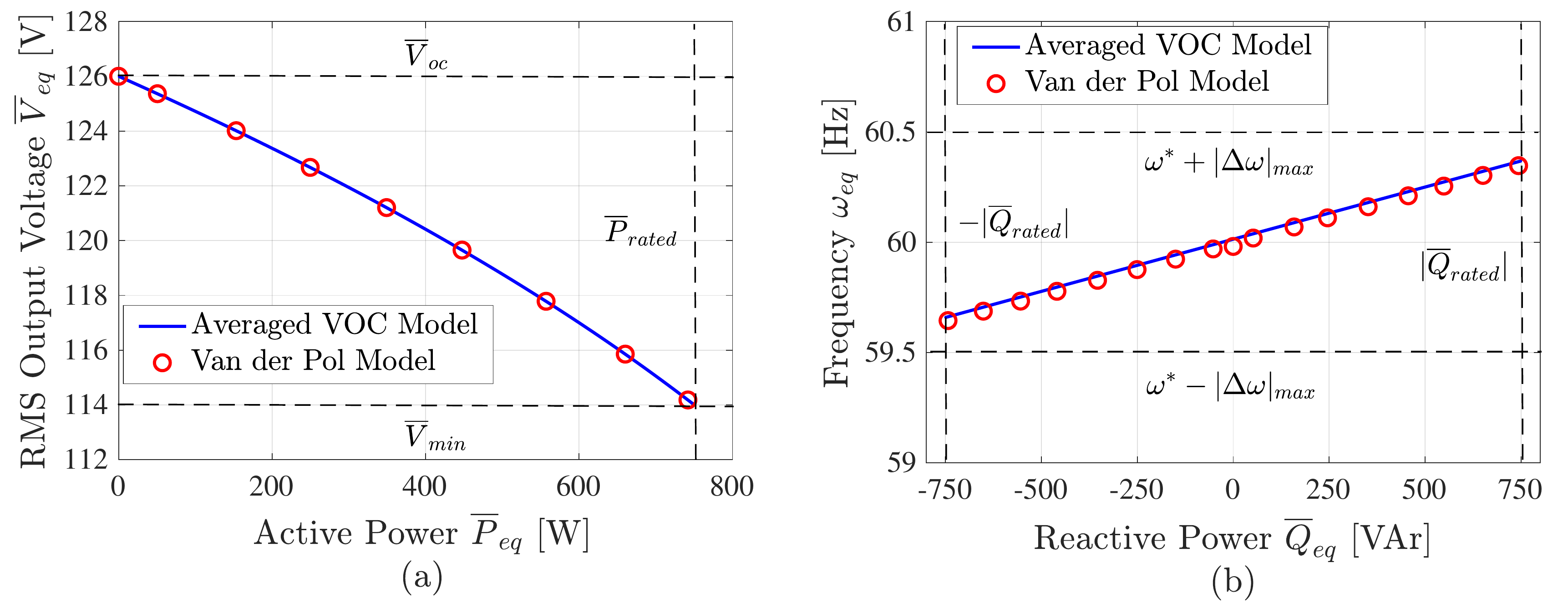}\\%{drivingpeaksrsq}
\vspace{-0.3 cm}
    \caption{A comparison between the embedded droop characteristics within the averaged VOC dynamics \eqref{eq:voc actual model v}-\eqref{eq:voc actual model t} and actual VOC dynamics \eqref{eq:voc actual model}.}
\label{fig:droopcharacteristics}
\vspace{-0.3 cm}
\end{figure}

In order to validate the embedded droop characteristics \eqref{eq:equil V1}, \eqref{eq:equil V4} within the averaged VOC dynamics, a comparison is made with the actual VOC dynamics \eqref{eq:voc actual model} with current feedback after the LCL filter. In Fig. \ref{fig:droopcharacteristics}, it can be seen that the embedded droop characteristics are close to the actual VOC dynamics. In order to obtain the $\overline{V}-\overline{P}$ and $\omega-\overline{Q}$ droop characteristics, the LCL filter is assumed to be ideal (i.e. $R_f =R_c = R_g = 0$) resulting in $S_{\alpha} = 0$ and the parameters in Table \ref{tab:System Parameters} are re-derived according to the design procedure in Section \ref{sec:VOC Parameters Design Procedure}. 
\vspace{-0.08 cm}
\subsection{Rise Time and Harmonics Analysis}
\label{subsec:Rise Time and Harmonics Analysis}
\begin{figure}[t]
\centering
\includegraphics[width=0.9\columnwidth]{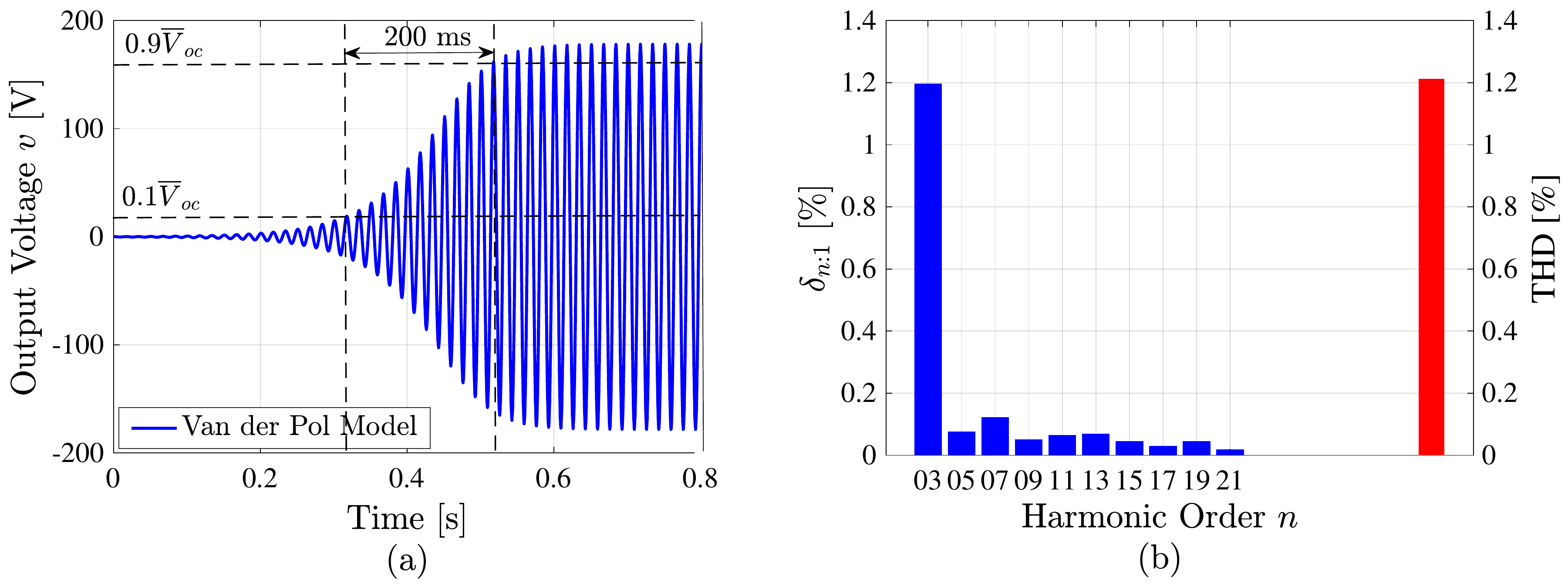}\\%{drivingpeaksrsq}
\vspace{-0.3 cm}
    \caption{The rise time and harmonics analysis shows that the virtual oscillator controlled inverter closely follows the desired design inputs $t_{rise}^{max}$ and $\delta_{3:1}^{max}$.}
\label{fig:risetimeharmonics}
\vspace{-0.5 cm}
\end{figure}

In Fig. \ref{fig:risetimeharmonics}, the rise time and harmonics analysis is presented for an unloaded inverter. It can be seen that the VO-controlled inverter closely follows the design inputs $t_{rise}^{max}$ and $\delta_{3:1}^{max}$, thus validating the parameter design procedure in Section \ref{sec:VOC Parameters Design Procedure}. 
%
%\vspace{-0.5 cm}
\subsection{Models Comparison}
\label{subsec:Models Comparison}
\begin{figure}[t]
\centering
\includegraphics[width=0.9\columnwidth]{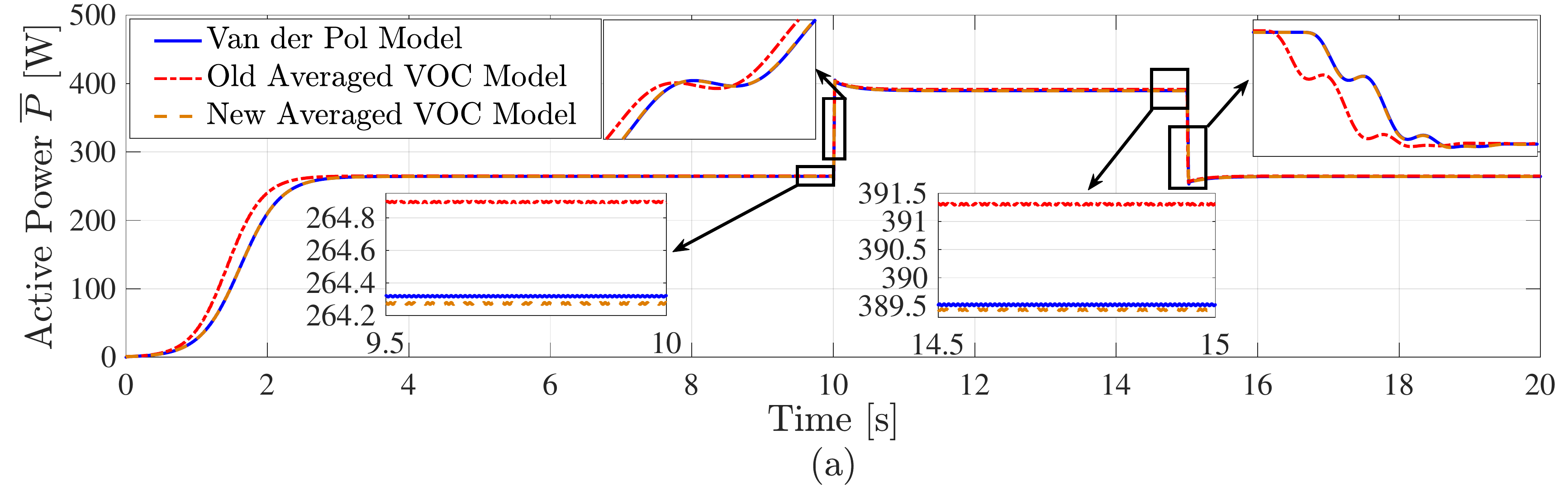}\\%{drivingpeaksrsq}
\includegraphics[width=0.9\columnwidth]{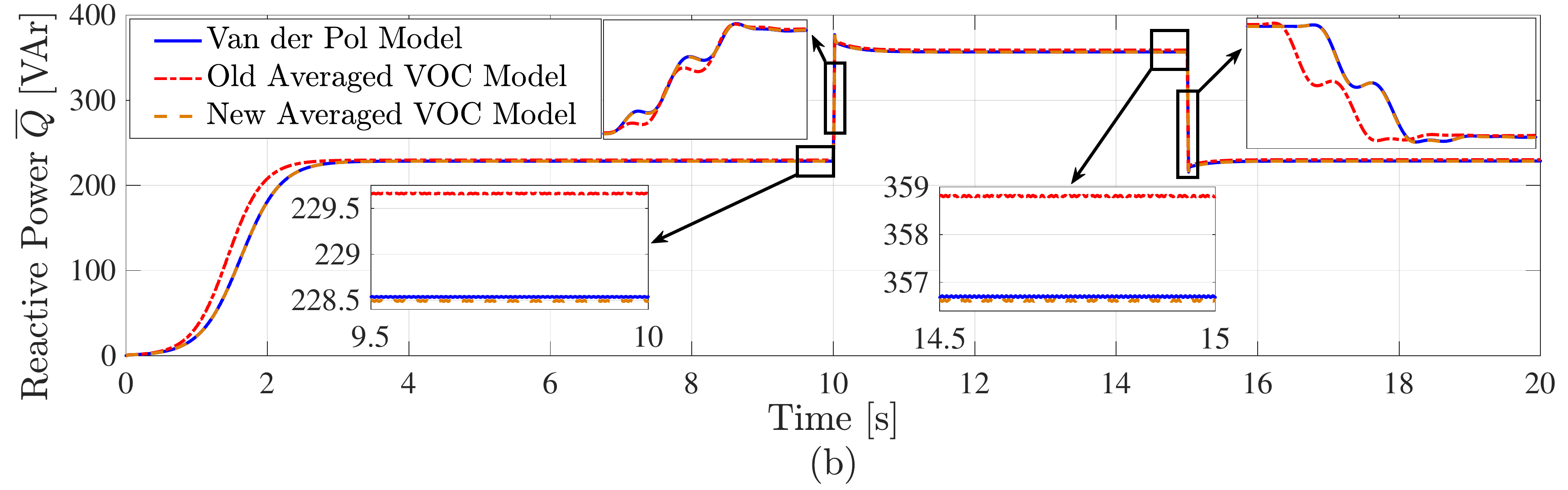}\\%{drivingpeaksrsq}
\includegraphics[width=0.9\columnwidth]{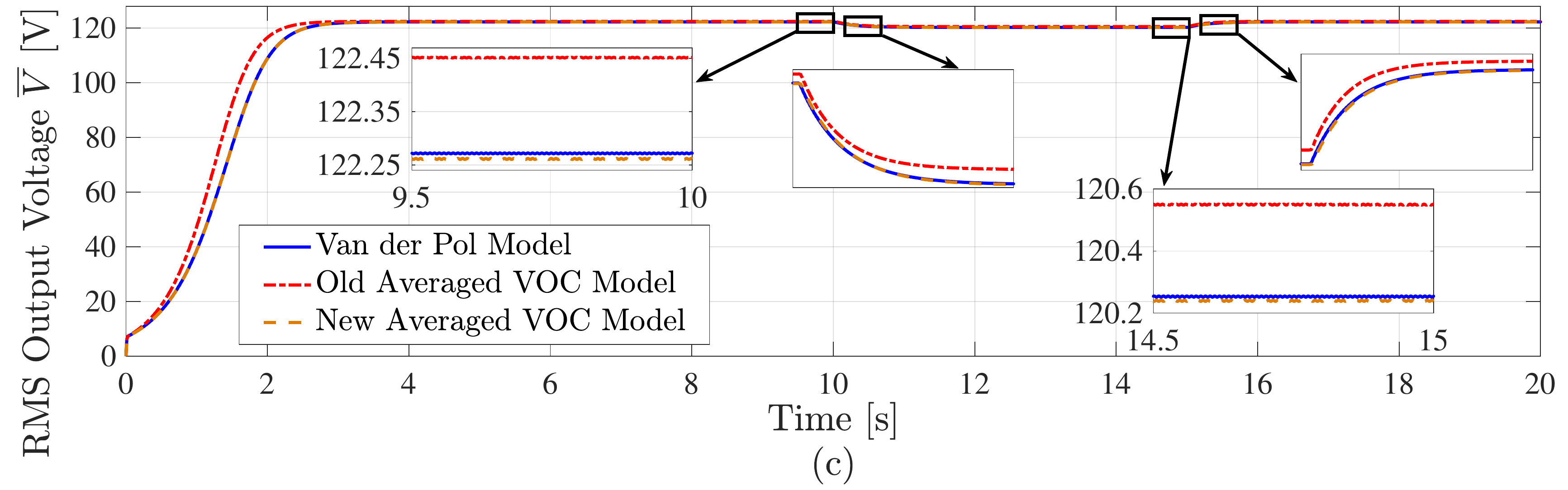}\\%{drivingpeaksrsq}
\includegraphics[width=0.9\columnwidth]{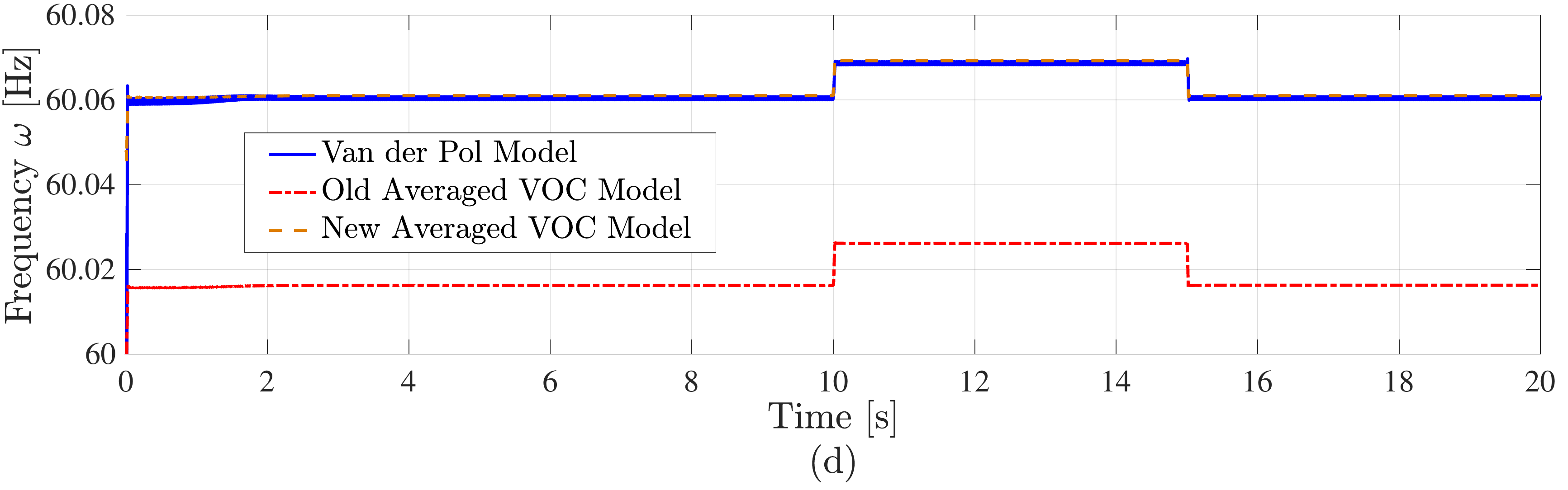}\\%{drivingpeaksrsq}
\vspace{-0.4 cm}
    \caption{Comparison between previously reported averaged VOC model \cite{johnson2016synthesizing}, the proposed averaged VOC model \eqref{eq:voc actual model v}-\eqref{eq:voc actual model t} and actual VOC dynamics \eqref{eq:voc actual model}: (a) active power, (b) reactive power, (c) voltages, (d) frequency.}
\label{fig:modelcomparison}
\vspace{-0.4 cm}
\end{figure}

In Fig. \ref{fig:modelcomparison}, a comparison is made between the previously reported averaged VOC model \cite{johnson2016synthesizing}, the proposed averaged VOC model \eqref{eq:voc actual model v}-\eqref{eq:voc actual model t} and actual VOC dynamics \eqref{eq:voc actual model}. Note that the proposed averaged VOC model follows the actual VOC dynamics more closely as compared to the previously reported averaged VOC model. For comparison, the LCL filter is chosen with higher values of filter parameters ($L_f^{\iota} = 10 L_f, C_f^{\iota} = 19.75 C_f$) and the epsilon is chosen to be $\epsilon^{\iota} = \frac{\epsilon}{8} = \frac{\sqrt{L/C}}{8}$. The $z_L = 6.9 +j16.6\,\Omega$ with $z_S = 44.24 + j10.85\,\Omega$ in-parallel to implement the step-up and step-down changes. The system parameters in Table \ref{tab:System Parameters} are re-derived according to the design procedure in Section \ref{sec:VOC Parameters Design Procedure}. Note that the difference between the previously reported averaged VOC model and actual VOC dynamics starts to increase for a higher value of filter capacitance. The output power is computed using a moving average window where window's length is updated based on the frequency recovered from actual VOC dynamics.%A large filter capacitor draws more current and makes the difference between currents before and after the LCL filter more significant.  
%
%\vspace{-0.1 cm}
\subsection{Power Dispatch}
\label{subsec:Power Dispatch}
\begin{figure}[t]
\centering
\includegraphics[width=0.9\columnwidth]{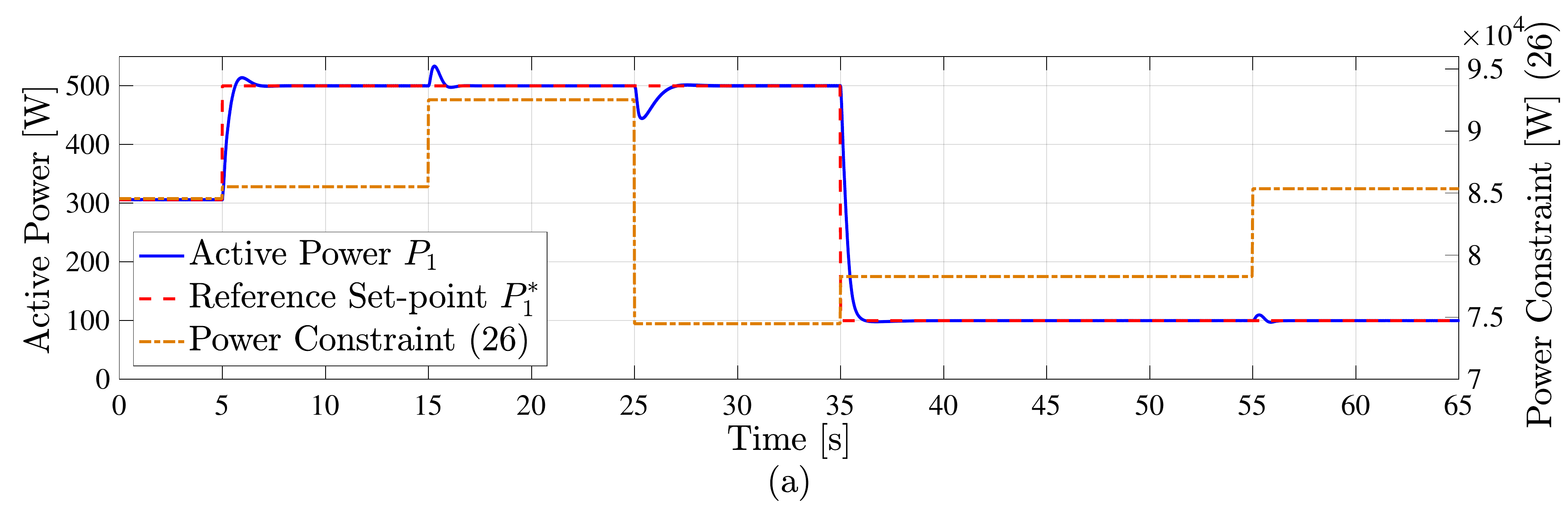}\\%{drivingpeaksrsq}
\includegraphics[width=0.9\columnwidth]{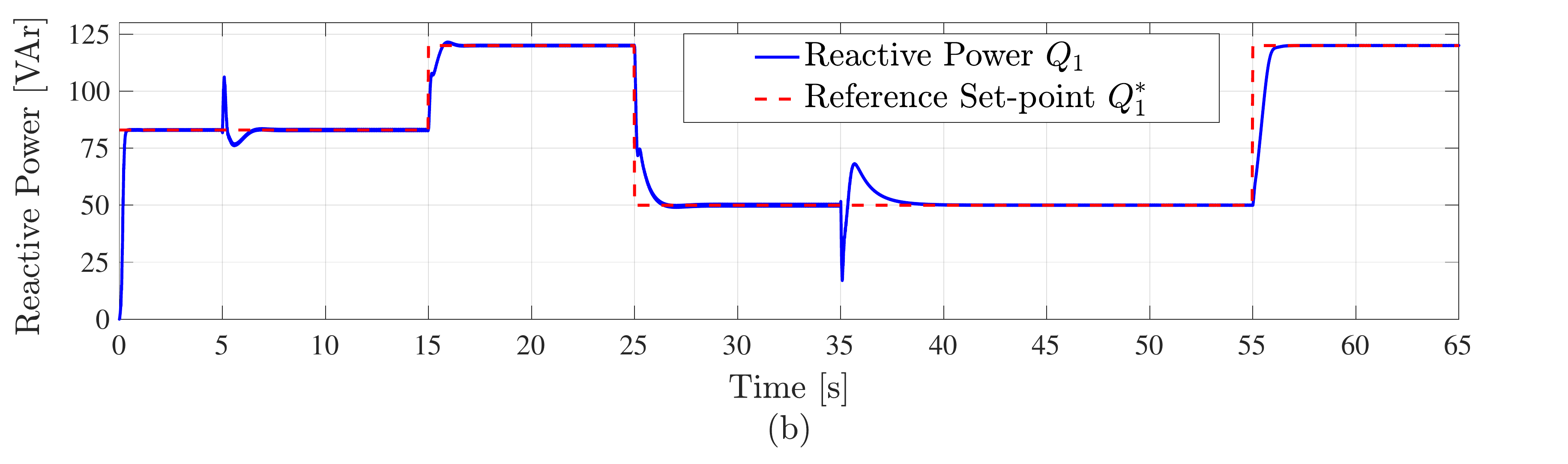}\\%{drivingpeaksrsq}
\includegraphics[width=0.9\columnwidth]{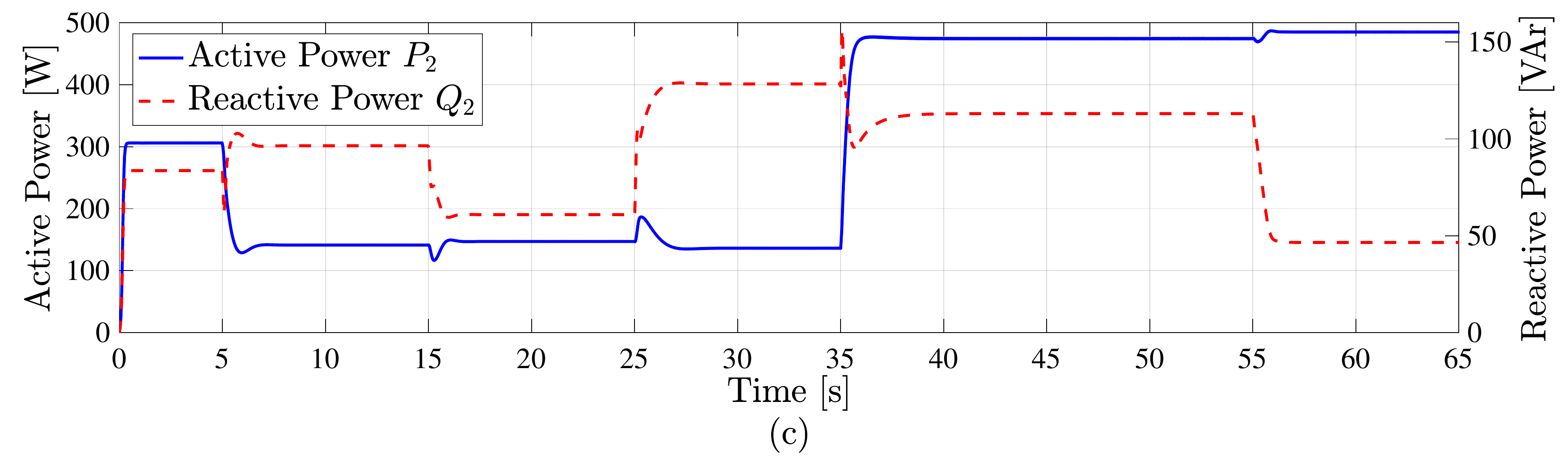}\\%{drivingpeaksrsq}
\includegraphics[width=0.9\columnwidth]{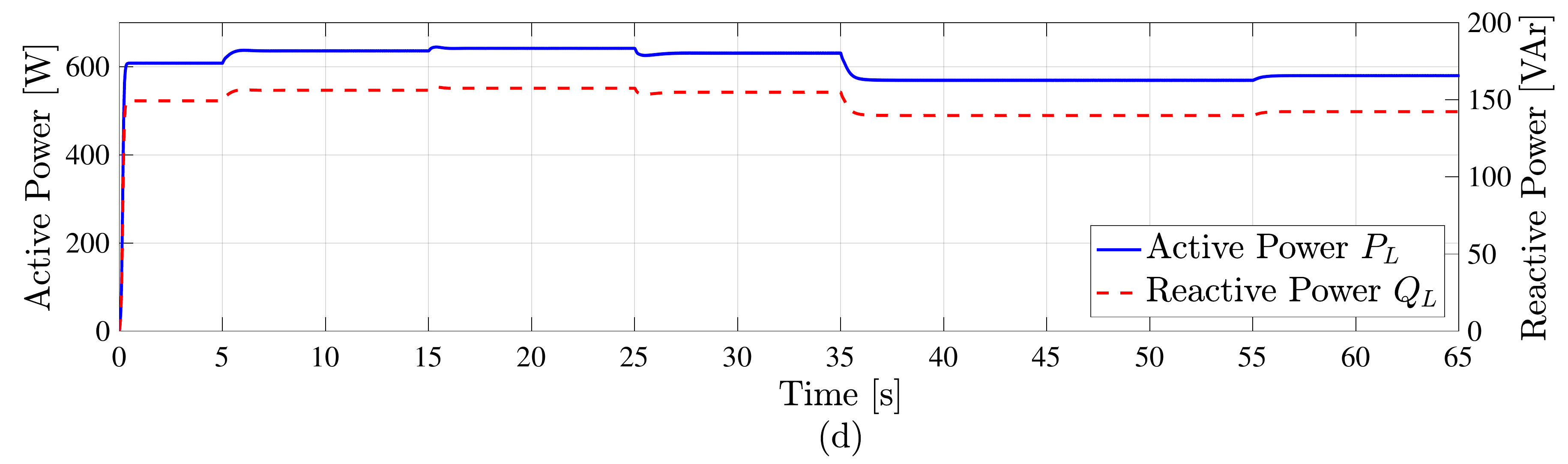}\\%{drivingpeaksrsq}
\includegraphics[width=0.9\columnwidth]{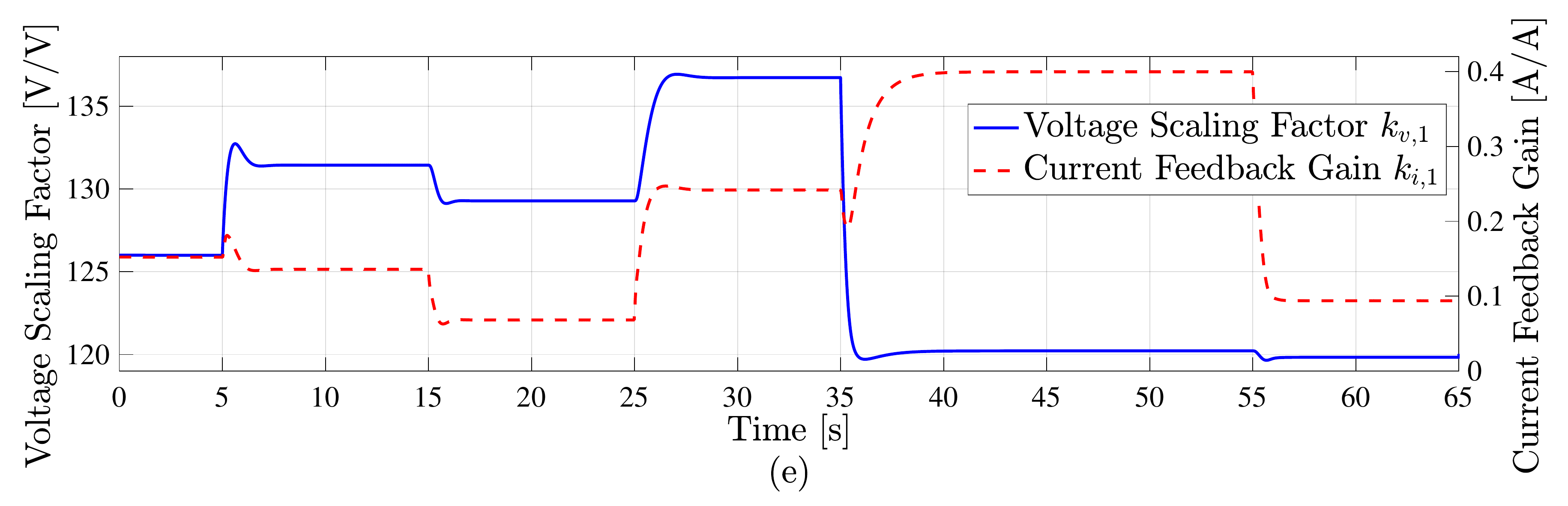}\\%{drivingpeaksrsq}
\includegraphics[width=0.9\columnwidth]{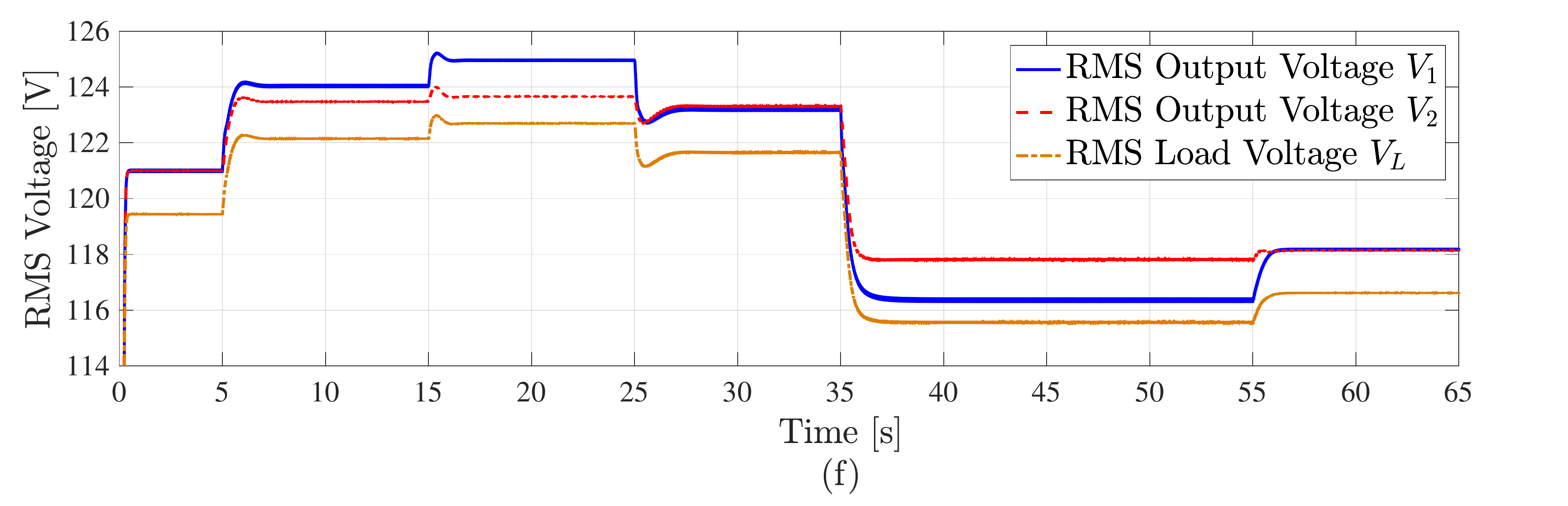}\\%{drivingpeaksrsq}
\vspace{-0.4 cm}
    \caption{Power dispatch results for inverter $1$ while inverter $2$ supplies the remaining load power: (a) inverter $1$ active power, power security constraint and reference set-point, (b) inverter $1$ reactive power and reference set-point, (c) inverter $2$ output power, (d) load power, (e) control inputs, (f) voltages.}
\label{fig:powerdispatch}
\vspace{-0.4 cm}
\end{figure}

In Fig. \ref{fig:powerdispatch}, the power dispatch results are presented for a number of scenarios as listed in Table \ref{tab:Power Dispatch Set-points}. Inverter $1$ simultaneously regulates both the active and reactive power using the two PI controllers \cite{ali2018simultaneous} that continuously tune the voltage scaling factor $k_{v,1}$ and current feedback gain $k_{i,1}$. The inverter $2$ supplies the remaining power to the load acting like a slack bus. The PI controllers' proportional and integral gains are $K_P^p = -0.001$, $K_I^p = -0.15$, $K_P^q = 0.0001$ and $K_I^q = 0.01$. In Fig. \ref{fig:powerdispatch}, inverter $1$ tracks the desired power set-points effectively while satisfying the security constraint \eqref{eq:constraint}. The corresponding changes in the inverter $2$ power, load power, control inputs and voltages are also presented.\\
\vspace{-0.0 cm}
\begin{table}[t]
\vspace{-0.0 cm}
\centering
\renewcommand{\arraystretch}{1.3}
\caption{Power Dispatch Set-points}
\label{tab:Power Dispatch Set-points}
%\tiny
\resizebox{0.85\columnwidth}{!}{
\begin{tabular}{>{\centering}p{1.5cm}>{\centering}p{1.5cm}>{\centering}p{1.5cm}>{\centering\arraybackslash}p{1.5cm}}
    \hline
    \hline
        Case No. & $P_1^*$ [W] & $Q_1^*$ [VAr] & Time [s] \\
    \hline
        $1$ & $-$ & $-$ & $0-5$ \\
        $2$ & $500$ & $83$ & $5-15$ \\
        $3$ & $500$ & $120$ & $15-25$ \\
        $4$ & $500$ & $50$ & $25-35$ \\
        $5$ & $100$ & $50$ & $35-55$ \\
        $6$ & $100$ & $120$ & $55-65$ \\
    \hline
    \hline
\end{tabular}}
\vspace{-0.55 cm}
\end{table}
\vspace{-0.5 cm}
\section{Conclusion}
\label{sec:Conclusion}
The LCL filter is considered an essential part of commercial inverters to filter out the harmonics and improve the output voltage. Keeping this in view, an averaged VOC model is derived for inverters with current feedback after the output LCL filter. The averaged model uncovers the embedded droop-characteristics within the averaged VOC dynamics and simplifies the analysis. The voltage and frequency regulation characteristics are determined. Further, to enable the VO-controlled inverter to satisfy the desired ac-performance specifications, a parameter design procedure is presented. Moreover, a recently reported power dispatch technique is extended to the proposed averaged VOC model with current feedback after the output LCL filter. The updated control laws are derived to determine the control inputs corresponding to a particular power set-point. An updated power security constraint is derived to determine the feasible operating region. The power security constraint is helpful in planning the optimal power flow in an electric grid. The simulation results validate the proposed averaged VOC model and power dispatch technique. Future work can include generalising the method to multiple inverters, stability analysis and experimental validation. 

\appendix
\subsection{Derivation of the Averaged VOC Model with an LCL Filter}
\label{subsec:Derivation of the Averaged VOC Model with an LCL Filter}
Consider an inverter with current feedback after the output LCL filter as shown in Fig. \ref{fig:twovocsystem}. Let us denote a phasor associated with a time domain variable $(.)$ by $\overrightarrow{(.)}$. The $\overrightarrow{V}$ denotes the voltage before the LCL filter, the $\overrightarrow{V_{o}}$ denotes the filter capacitor voltage and the $\overrightarrow{V_g}$ denotes the voltage after the LCL filter. Similarly, the $\overrightarrow{I_f}$ denotes the current flowing through filter inductor $L_f$, the $\overrightarrow{I_{c}}$ denotes the filter capacitor current and the $\overrightarrow{I_g}$ is the current flowing through filter inductor $L_g$. Solving the network equations, we have:
\footnotesize
\begin{align}\label{eq:AV22}
    \overrightarrow{I_g} = \left[ \begin{array}{cc} Z_{\alpha} \angle \theta_{\alpha} & Z_{\beta} \angle \theta_{\beta} \end{array} \right] \left[ \begin{array}{c} \overrightarrow{I_f} \\ \overrightarrow{V} \end{array} \right],
\end{align}
\normalsize 
where $z_{\alpha} = Z_{\alpha} \angle \theta_{\alpha} = \frac{z_c + z_f}{z_c}$ and $z_{\beta} = Z_{\beta} \angle \theta_{\beta} = \frac{-1}{z_c}$ are the impedance constants defined at $\omega^*$. Further, we have \cite{johnson2016synthesizing}:
\footnotesize
\begin{align}\label{eq:AV23}
    \frac{d}{dt} \phi = \omega^* + \frac{d}{dt} \theta^* = \omega + \frac{d}{dt} \theta,
\end{align}
\normalsize
where the instantaneous phase angle is denoted by $\phi$, the nominal grid frequency is denoted by $\omega^*$ and the load dependent steady-state frequency is denoted by $\omega$. The angles $\theta^*$ and $\theta$ denote the phase offset with respect to $\omega^*t$ and $\omega t$, respectively. The inverter output voltage is given by:
\footnotesize
\begin{align}\label{eq:AV233}
    v(t) = \sqrt{2} V(t) \cos{(\omega^* t + \theta^*(t))}.
\end{align}
\normalsize
The instantaneous active and reactive output power of inverter in terms of $v(t)$ and $i(t)$ is given by:
\footnotesize
\begin{align}\label{eq:AV24}
    P(t) = v(t)i(t), \quad \quad Q(t) = v\left( t - \frac{\pi}{2}\right) i(t).
\end{align}
\normalsize 
The averaged active and reactive power over an ac-cycle $\frac{2\pi}{\omega^*}$ is defined as:
\footnotesize
\begin{align}\label{eq:AV25}
    \overline{P}(t) = \frac{\omega^*}{2\pi} \int_{s = t}^{t+2\pi/\omega^*} P(s)ds, \quad \overline{Q}(t) = \frac{\omega^*}{2\pi} \int_{s = t}^{t+2\pi/\omega^*} Q(s)ds.
\end{align}
\normalsize 
In order to derive the averaged dynamics \eqref{eq:voc actual model v}-\eqref{eq:voc actual model t}, a change of variable is made from $t \rightarrow \tau = \omega^*t$ and $\theta^*(\tau) = \phi(\tau/\omega^*) -\tau$. Expressing the actual VOC dynamics \eqref{eq:voc actual model} as a function of $\theta^*$, we have \cite{johnson2016synthesizing}:
\footnotesize
\begin{align}\label{eq:AV26}
    \notag
    \frac{dV}{d\tau} &= \frac{\epsilon}{\sqrt{2}} \left( \sigma g\big(\sqrt{2}V\cos{(\tau + \theta^*)}\big) - k_{v}k_{i}i\right)\cos{(\tau + \theta^*)},\\
    \frac{d\theta^*}{d\tau} &= - \frac{\epsilon}{\sqrt{2}V}\left(\sigma g\big(\sqrt{2}V\cos{(\tau + \theta^*)}\big)-k_{v}k_{i}i\right)\sin{(\tau + \theta^*)}.
\end{align}
\normalsize
The dynamics in \eqref{eq:AV26} are $2\pi$ periodic functions in $\tau$. The averaged dynamics in the quasi-harmonic limit $\epsilon \searrow 0$ (following \cite[Eq. 12]{johnson2016synthesizing}) are given by:
\footnotesize
\begin{align}\label{eq:AV27}
    \notag
    \left[\begin{array}{c}\dot{\overline{V}}\\\dot{\overline{\theta}}^*\end{array} \right]=&\frac{\epsilon\sigma}{2\pi\sqrt{2}}\int_{0}^{2\pi} g\left(\sqrt{2}\overline{V} \cos{(\tau+\overline{\theta}^*)}\right)\left[\begin{array}{c}\cos{(\tau+\overline{\theta}^*)}\\\frac{-1}{\overline{V}}\sin{(\tau+\overline{\theta}^*)}\end{array}\right] d\tau\\
    \notag
    &-\frac{\epsilon k_vk_i}{2\pi\sqrt{2}}\int_{0}^{2\pi}i\left[\begin{array}{c}\cos{(\tau+\overline{\theta}^*)}\\ \frac{-1}{\overline{V}}\sin{(\tau+\overline{\theta}^*)} \end{array}\right]d\tau,\\  
    =&\frac{\epsilon\sigma}{2}\left[\begin{array}{c}\overline{V} - \frac{\beta}{2}\overline{V}^3 \\ 0 \end{array}\right] -\frac{\epsilon k_vk_i}{2\pi\sqrt{2}}\int_{0}^{2\pi}i\left[\begin{array}{c}\cos{(\tau+\overline{\theta}^*)}\\ \frac{-1}{\overline{V}}\sin{(\tau+\overline{\theta}^*)} \end{array}\right]d\tau.
\end{align}
\normalsize
Changing the coordinates from $\tau$ to $t$ in \eqref{eq:AV27} and keeping the $\mathcal{O}(\epsilon)$ terms only, we have:
\footnotesize
\begin{align}\label{eq:AV277}
    \notag
    \frac{d}{dt}\left[\begin{array}{c}\overline{V} \\ \overline{\theta}^*\end{array} \right] &= \frac{\sigma}{2C} \left[\begin{array}{c}\overline{V} - \frac{\beta}{2}\overline{V}^3 \\ 0 \end{array}\right] \\
    &-\frac{k_vk_i \omega^*}{2\pi\sqrt{2}C}\int_{0}^{\frac{2\pi}{\omega^*}}i(t)\left[\begin{array}{c}\cos{(\omega^*t+\overline{\theta}^*)}\\ \frac{-1}{\overline{V}}\sin{(\omega^*t+\overline{\theta}^*)} \end{array}\right]dt.  
\end{align}
\normalsize
Lets define the impedance constants $C_{\alpha} = Z_{\alpha}\cos{\theta_{\alpha}}$, $C_{\beta} = Z_{\beta}\cos{\theta_{\beta}}$, $S_{\alpha} = Z_{\alpha}\sin{\theta_{\alpha}}$ and $S_{\beta} = Z_{\beta}\sin{\theta_{\beta}}$. The current $i = \sqrt{2}I\cos{(\omega^*t + \theta_i^*)}$ where $I$ is the RMS current magnitude and $\theta_i^*$ is the phase-offset with respect to $\omega^*t$. Let $\overrightarrow{I_g}^{\Re}$ denotes the real part of the current $\overrightarrow{I_g}$ \eqref{eq:AV22} and is given by:
\footnotesize
\begin{align}\label{eq:AV28}
%\notag
    \overrightarrow{I_g}^{\Re} = \sqrt{2}\left(Z_{\alpha}I\cos{(\omega^*t + \theta^*_i + \theta_{\alpha})} + Z_{\beta}V\cos{(\omega^*t + \theta^* + \theta_{\beta})}\right). 
\end{align}
\normalsize
In order to derive the averaged VOC model, the current $i = i_g$ is replaced by its real part $\overrightarrow{I_g}^{\Re}$ in \eqref{eq:AV277}, we get:
\footnotesize
\begin{align}\label{eq:AV29}
    \notag
    \frac{d}{dt}\left[\begin{array}{c}\overline{V} \\ \overline{\theta}^*\end{array} \right] & = \frac{\sigma}{2C} \left[\begin{array}{c}\overline{V} - \frac{\beta}{2}\overline{V}^3 \\ 0 \end{array}\right] -\frac{k_vk_i \omega^*}{4\pi C}\int_{0}^{\frac{2\pi}{\omega^*}} \left[ \begin{array}{c} \vphantom{} \\ \vphantom{} \end{array} \right. \\ 
    \notag
    & \left. \begin{array}{c} \frac{Z_{\alpha}}{\overline{V}} \sqrt{2}V(t)\cos{(\omega^*t+\theta^*)} \sqrt{2}I(t)\cos{(\omega^*t + \theta^*_i + \theta_{\alpha})} + \\ \frac{-Z_{\alpha}}{\overline{V}^2} \sqrt{2}V(t)\sin{(\omega^*t+\theta^*)} \sqrt{2}I(t)\cos{(\omega^*t + \theta^*_i + \theta_{\alpha})} - \end{array} \right.\\
 & \left. \begin{array}{c} 2Z_{\beta}\overline{V}\cos{(\omega^*t + \theta^* + \theta_{\beta})} \cos{(\omega^*t + \theta^*)} \\ 2Z_{\beta}\cos{(\omega^*t + \theta^* + \theta_{\beta})} \sin{(\omega^*t + \theta^*)} \end{array} \right]dt.    
\end{align}
\normalsize
Using the trigonometric identities and definition of instantaneous active and reactive power \eqref{eq:AV24}, we get:
\footnotesize
\begin{align}\label{eq:AV30}
    \notag
    &\frac{d}{dt}\left[\begin{array}{c}\overline{V} \\ \overline{\theta}^*\end{array} \right]  = \frac{\sigma}{2C} \left[\begin{array}{c}\overline{V} - \frac{\beta}{2}\overline{V}^3 \\ 0 \end{array}\right] -\frac{k_vk_i \omega^*}{4\pi C}\int_{0}^{\frac{2\pi}{\omega^*}} \left[ \begin{array}{c} \vphantom{} \\ \vphantom{} \end{array} \right. \\ 
     &\left. \begin{array}{c} \frac{1}{\overline{V}} \left( C_{\alpha}P(t) + S_{\alpha}Q(t) \right) + Z_{\beta}\overline{V} \left ( \cos{\theta_{\beta} + \cos{(2\omega^*t + 2\theta^* + \theta_{\beta})}} \right) \\ \frac{-1}{\overline{V}^2} \left( C_{\alpha}Q(t) - S_{\alpha}P(t)\right ) - Z_{\beta} \left( \sin{(2\omega^*t + 2\theta^* + \theta_{\beta})} - \sin{\theta_{\beta}} \right) \end{array} \right]dt.    
\end{align}
\normalsize
The averaged VOC dynamics \eqref{eq:voc actual model v}-\eqref{eq:voc actual model t} are recovered from \eqref{eq:AV30}.
\section*{Acknowledgment}
The authors acknowledge the kind support of A.W. Tyree Foundation, and Australian Research Council through Discovery Projects Funding Scheme under Project DP180103200.
\bibliographystyle{ieeetr}
\bibliography{cpsreferences}
\end{document}